\documentstyle[12pt]{article}
\begin{document}
\renewcommand{\theequation}{\mbox{\arabic{section}.\arabic{equation}}}
\def\t{\times}\def\p{\phi}\def\P{\Phi}\def\a{\alpha}
\def\e{\varepsilon}\def\be{\begin{equation}}\def\ee{\end{equation}}
\def\l{\label}\def\0{\setcounter{equation}{0}}\def\b{\beta}\def\S{\Sigma}
\def\C{\cite}\def\r{\ref}\def\ba{\begin{eqnarray}}\def\ea{\end{eqnarray}}
\def\n{\nonumber}\def\R{\rho}\def\X{\Xi}\def\x{\xi}\def\La{\Lambda}
\def\la{\lambda}\def\d{\delta}\def\s{\sigma}\def\f{\frac}\def\D{\Delta}
\def\pa{\partial}\def\Th{\Theta}\def\o{\omega}\def\O{\Omega}\def\th{\theta}
\def\ga{\gamma}\def\Ga{\Gamma}\def\h{\hat}\def\rar{\rightarrow}
\def\vp{\varphi}\def\inf{\infty}\def\le{\left}\def\ri{\right}
\def\foot{\footnote}\def\u{\underline}

\begin{center}

{\Large\bf Dual-Resonance Model and Multiple Production}
\vskip 0.3cm
{\large\it J.Manjavidze, A.Sissakian, H.Torosian\\ {JINR, Dubna, Russia}}

\end{center}

\begin{abstract}
The very high multiplicity (VHM) processes are considered to
investigate consequences of the dual resonance models reach
(exponential) mass spectrum.  The virial decomposition is developed
for description of the produced particles short-rang (resonance)
correlations. It is shown that the reach mass spectrum is able to
do more flat the VHM distribution: $\s_n=O(e^{-n})$ up to
multiplicities $n\simeq\bar{n}^2$, $\bar{n}$ is the mean
multiplicity.  But for $n>>\bar{n}^2$ one should expect
$\s_n<O(e^{-n})$.
\end{abstract}

\section{Introduction}

The last two decades of field theory development is marked by
considerable efforts to avoid the problem of color charge confinement
formulating a closed hadron field theory. The remarkable attempt
based on the string model, in its various realizations \C{string}.
But, in spite of remarkable success (in formalism especially) there
is not an experimentally measurable predictions of this approach till
now, e.g.  \C{str.-exp.}.

The string model is a natural consequence of the old dual resonance
model \C{dual} and we hope that our toy approach includes main
characteristic features of this model. We would like to describe in
this paper production of `stable' hadrons through decay of
resonances.  This channel was considered firstly in the papers
\C{tor}.

Our consideration will use following assumptions.

{\bf A}. The string interpretation of the dual-resonance model bring
to the observation that the mass spectrum of resonances, i.e.  the
total number $\R(m)$ of mass $m$ resonance excitations, grows
exponentially:
\be
\R(m)=(m/m_0)^{\ga} e^{\b_0m},~\b_0={\rm const},~m>m_0.
\l{rms}\ee
Note also that the same hadron mass spectrum (\r{rms}) was predicted
in the `bootstrap' approach \C{hag}.  It predicts that
\be
\ga=-5/2.
\l{crex}\ee

Another assumptions are based on the ordinary (resonance
$\leftrightarrow$ Regge pole) duality.

{\bf B}. The `probability' of mass $m$ resonance
creation $\s^R(m)$ has the Regge pole asymptotics:
\be
\s^R(m)=g^R\f{m_0}{m},~m\geq m_0\approx 0.2 {\rm Gev},~g^R={\rm
const.}
\l{rcs}\ee
It was assumed here that the intercept of Regge pole trajectory
$\a^R=1/2$. So, the meson resonances only would be taken into
account.

{\bf C}. If $\s^R_n(m)$ describes decay of mass $m$ resonance on the
$n$ hadrons, then the mean multiplicity of hadrons
\be
\bar{n}^R(m)=\f{\sum_n n\s^R_n(m)}{\s^R(m)}.
\l{rmm}\ee
Following to the Regge model,
\be
\bar{n}^R(m)=\bar{n}^R_0\ln\f{m^2}{m_0^2}.
\l{rmm2}\ee

{\bf D}. We will assume that there is a definite vicinity of
$\bar{n}^R(m)$ where $\s^R_n(m)$ is defined by $\bar{n}^R(m)$ only.
I.e. in this vicinity
\be
\s^R_n(m)=\s^R(m)e^{-\bar{n}^R(m)}({{\bar{n}^R(m)})^n}/{n!}.
\l{pois}\ee
This is the direct consequence of the Regge pole model, if $m/m_0$ is
high enough.

The connection between $S$-matrix approach and the real-time
statistics (finite temperature field theories) \C{elpat} will be used
to formulate our model quantitatively. This interpretation will be
useful since it allows to formulate the description in terms of a few
parameters only. All this statistical parameters are expressed
through created particles energies and momenta.

We will use also the virial decomposition technique. It was extremely
effective for description of the phase transitions critical region,
where the correlation radii tends to infinity. The Mayer'
decomposition over `connected groups' is well known in this
connection \C{mayer}.

Following to our idea, we will distinguish the short-range
correlations among hadrons and the long-range correlations among
resonances. The `connected groups' would be described by resonances
(strings) and the interactions among them should be described
introducing for this purpose the correlation functions among strings.
So, we will consider the `two-level' model of hadrons creation: first
level describes the short-range correlation among hadrons and the
second level is connected to the correlations among strings.

\section{Asymptotic estimations}

Our purpose is to investigate the role of exponential spectrum
(\r{rms}) in the asymptotics over multiplicity $n$. In this case one
can valid heavy resonances creation and such formulation of the
problem have definite advantage.

(i) If creation of heavy resonances at $n\to\infty$ is expected, then
one can neglect the dependence on resonances momentum ${\bf q_i}$ .
So, the `low-temperature' expansion is valid in the VHM region.

(ii) Having the big parameter $n$ one can construct the perturbations
expanding over $1/n$. We will see that there is a wide domain for
$n$, where one can neglect resonance correlations.

(iii) We will be able to show at the end the range of applicability of
this assumptions. For this purpose following formal phenomena will be
used. Let us introduce the `grand partition function'
\be
\X(z,s)=\sum_n z^n \s_n(s),~\X(1,s)=\s_{tot}(s),~
n\leq\sqrt{s}/m_0\equiv n_{max}(s),
\l{gpf}\ee
and let us assume that just $\X$ is known. Then, using the inverse
Mellin transformation,
\be
\s_n(s)=\f{1}{2\pi i}\int\f{dz}{z^{n+1}}\X(z,s).
\l{metr}\ee
This integral will be computed expanding it in vicinity of solution
$z_c>0$ of equation:
\be
n=z\f{\pa}{\pa z}\ln\X(z,s).
\l{eqst1}\ee
It is assumed, and this should be confirmed at the end, that the
fluctuations in vicinity of $z_c$ are Gaussian.

It is natural at first glance to consider $z_c=z_c(n,s)$ as the
increasing function of $n$. Indeed, this immediately follows from
positivity of $\s_n(s)$ and finiteness of $n_{max}(s)$ at finite $s$.
But one can consider the limit $m_0\to0$.  Theoretically this limit
is rightful because of PCAC hypotheses and nothing should be
happen if the pion mass $m_0\to0$.  In this sense $\X(z,s)$ may be
considered as the whole function of $z$.  Then, $z_c=z_c(n,s)$ would
be increasing function of $n$ if and only if $\X(z,s)$ is regular
function at $z=1$.

The prove of this statement is as follows. We should conclude, as
follows from eq.(\r{eqst1}), that
\be
z_c(n,s)\to z_s~{\rm at}~n\to\infty,~{\rm and~at}~s=const,
\l{asy}\ee
i.e. the singularity points $z_s$ $attracts$ $z_c$ in asymptotics
over $n$.

If $z_s=1$, then $(z_c-z_s)\to +0$, when $n$ tends to infinity
\C{sys}.  But if $z_s>1$, then $(z_c-z_s)\to-0 $ in VHM region.

On may find the estimation:
\be
-\f{1}{n}\ln\f{\s_n(s)}{\s_{tot}(s)}= \ln z_c(n,s)+O(1/n),
\l{est0}\ee
where $z_c$ is the $smallest$ solution of (\r{eqst1}). It should be
underlined that this estimation is $independent$ on the character of
singularity, i.e. the position $z_s$ only is important.

\section{ Partition function}

Introducing the `grand partition function' (\r{gpf}) the `two-level'
description means that
\ba
\ln\f{\X(z,\b)}{\s_{tot}(s)}=-\b{\cal F}(z,s)=
\n\\
=\sum_k\f{1}{k!}\int
\prod^k_{i=1}\le\{\f{d^3q_idm_i\x(q_i,z)e^{-\b\e_i}}{(2\pi)^32\e_i}
\ri\}
N_k(q_1,q_2,...,q_k;\b),
\l{vd}\ea
where $\e_i=\sqrt{q_i^2+m_i^2}$. This is our virial decomposition.
Indeed, by definition
\be
\X(z,s)\le.\ri|_{\x=1}=\s_{tot}(s).
\l{tot}\ee
The quantity $\x(q,z)$ may be considered as the local activity. So,
\be
\le.\f{\d\X}{\d\x(q,z)}\ri|_{\x=1}\sim\s_{tot}N_1(q)
\l{r1cf}\ee
So, if decay of resonances form a group with 4-momentum $q$, then
$N_1(q)$ is the mean number of such groups. The second derivative
gives:
\be
\le.\f{\d^2\X}{\d\x(q_1,z)\d\x(q_2,z)}\ri|_{\x=1}\sim\s_{tot}
\{N_2(q_1,q_2)-N_1(q_1)N_1(q_2)\}\equiv\s_{tot}K_2(q_1,q_2)
\l{r2cf}\ee
where $K_2(q_1,q_2)$ is two groups correlation function, and so on.

The Lagrange multiplier $\b$ was introduced in (\r{vd}) to each
resonance: the Bolzmann exponent $\exp\{-\b\e\}$ takes into
account the energy conservation law $ \sum_i \e_i=E,$ where $E$ is
the total energy of colliding particles, $2E=\sqrt{s}$ in the CM
frame. This conservation law means that $\b$ is defined by equation:
\be
\sqrt{s}=\f{\pa}{\pa\b}\ln\X(z,\b).
\l{eqst2}\ee
So, to define the state one should solve two equations of state
(\r{eqst1}) and (\r{eqst2}).

The solution  $\b_c$ of the eq.(\r{eqst2}) have meaning of inverse
temperature of gas of strings if and only if the fluctuations in
vicinity of $\b_c$ are Gaussian.

On the second level we should describe the resonances decay onto
hadrons. Using (\r{pois}) we can write in some vicinity of $z=1$:
\be
\x(q,z)=\sum_n z^n \s^R_N(q)=g^R(\f{m_0}{m})e^{(z-1)\bar{n}(m)},
~m=|q|.
\l{rgpf}\ee
The assumptions {\bf B} and {\bf D} was used here.

So,
\be
-\b{\cal F}(z,s)=\sum_k\int
\prod^k_{i=1}\{dm_i^2\x(m_i,z)\}\tilde{N}_k(m_1,m_2,...,m_k;\b),
\l{vd1}\ee
where $\x$ was defined in (\r{rgpf}) and
\be
\tilde{N}_k(m_1,m_2,...,m_k;\b)=\int\prod_{i=1}^k\le\{
\f{d^3q_ie^{-\b\e_i(q_i)}}{2\e_i(q_i)}\ri\}
N_k(q_1,q_2,...,q_k;m_1,m_2,...,m_k).
\l{1}\ee
Assuming now that $|q_i|<<m$ are essential,
\be
\tilde{N}_k(m_1,m_2,...,m_k;\b)\simeq N'_k(m_1,m_2,...,m_k)
\prod_{i=1}^k\le\{\sqrt{\f{2m_i}{\b^3}}e^{-\b m_i}\ri\}
\l{est}\ee

Following to the duality assumption one may assume that
\be
N'_k(m_1,m_2,...,m_k)=\bar{N}_k(m_1,m_2,...,m_k)
\prod_{i=1}^k \le\{m_i^\ga e^{\b_0 m_i}\ri\}
\l{string}\ee
and $\bar{N}_k(m_1,m_2,...,m_k)$ is slowly varying function:
$$
\bar{N}_k(m_1,m_2,...,m_k)\simeq C_k
$$
In result the low-temperature expansion looks as follows:
\be
-\b{\cal F}(z,s)=\sum_k \f{2^{k/2}m_0^k(g^R)^kC_k}{\b^{3k/2}}
\le\{\int_{m_0}^\infty dm m^{\ga+3/2}
e^{(z-1)\bar{n}^R(m)-(\b-\b_0)m}\ri\}^k.
\l{fe1}\ee
We should assume that $(\b-\b_0)\geq0$. In this sense one may
consider $1/\b_0$ as the limiting temperature and above mentioned
constraint means that the string energies should be high enough.

\section{Thermodynamical parameters}

Remembering that the position of singularity over $z$  is essential
only, let us assume that the resonance interactions can not
renormalize it. Then, living first term in the sum (\r{fe1}),
\be
-\b{\cal F}(z,s)=\f{m_0g^RC_1}{\b^{3/2}}
\int_{m_0}^\infty dm (m/m_0)^{\ga+3/2}
e^{(z-1)\bar{n}^R(m)-(\b-\b_0)m}.
\l{fe2}\ee
We expect that this assumption is hold if
\be
n\to\infty,~s\to\infty,~\f{nm_0}{\sqrt{s}}\equiv\f{n}{n_{max}} <<1.
\l{restr}\ee

So, we would solve our equations of state with following `free
energy':
\be
-\b{\cal F}(z,s)=\f{\a}{\b^{3/2}}
\int_{m_0}^\infty d(\f{m}{m_0}) (\f{m}{m_0})^{\ga'-1}
e^{-\D(m/m_0)},
\l{fe3}\ee
where, using (\r{crex}),
\be
\ga'=\ga+2(z-1)\bar{n}^R_0+5/2=2(z-1)\bar{n}^R_0,~
\D=m_0(\b-\b_0)\geq0,~\a=const.
\l{nw}\ee

We have in terms of this new variables following equation for $z$,
\be
n=z\f{2\a\bar{n}^R_0}{\b^{3/2}}\f{\pa}{\pa\ga'}
\f{\Ga(\ga',\D)}{\D^{\ga'}}.
\l{eq1c}\ee
The equation for $\b$ takes the form:
\be
n_{max}=\f{\a m_0}{\b^{3/2}}\f{\Ga(\ga'+1,\D)}{\D^{\ga'+1}},
\l{eq2c}\ee
where $n_{max}=(\sqrt{s}/m_0)$ and $\Ga(\D,\ga')$ is the
incomplete $\Ga$-function:

$$
\Ga(\ga',\D)=\int^\infty_{\D} dx x^{\ga'-1}e^{-x}.
$$

\section{Asymptotic solutions}

Following to physical intuition one should expect the cooling of the
system when $n\to\infty$ (at fixed $\sqrt{s}$) and heating when
$n_{max}\to\infty$ (at fixed $n$).  But, as was mentioned above,
since the solution of eq.(\r{eq2c}) $\b_c$ is defined the value of
total energy, one should expect that $\b_c$ decrease in both cases.
So, the solution
\be
\D_c\geq0,~ \f{\pa\D_c}{\pa
n}<0~{\rm at}~n\to\infty,~ \f{\pa\D_c}{\pa s}<0~{\rm at}~s\to\infty
\l{sol2}\ee
is natural for our consideration.

The physical meaning of $z$ is activity. It defines at $\b=const$
the work needed for one particles creation. Then, if the system is
stable and $\X(z,s)$ may be singular at $z>1$ only,
\be
\f{\pa z_c}{\pa n}>0~{\rm at}~n\to\infty,~
\f{\pa z_c}{\pa s}<0~{\rm at}~s\to\infty.
\l{sol3}\ee

One should assume solving equations (\r{eq1c}) and (\r{eq2c}) they
\be
z_c\D^{\ga'_c+1}\f{\pa}{\pa\ga'_c}
\f{\Ga(\ga'_c,\D_c)}{\D_c^{\ga'_c}}<<\Ga(\ga'_c+1,\D_c).
\l{ineq1}\ee
This condition stands the physical requirement that $n<<n_{max}$. In
opposite case the finiteness of the phase space for $m_0\neq0$ should
be taken into account.

As was mentioned above the singularity $z_s$ attracts $z_c$ at
$n\to\infty$. By this reason one may consider following solutions.

A. $z_s=\infty$: $z_c>>\D$, $\D<<1$.

In this case
\be
\D^{-\ga'}\Ga(\ga',\D)\sim e^{\ga'\ln(\ga'/\D)}.
\l{a1}\ee
This estimation gives following equations:
\ba
n=C_1\ga'\ln(\ga'/\D)e^{\ga'\ln(\ga'/\D)},
\n\\
\f{n}{n_{max}}=C_2 \D\ga'\ln(\f{\ga'}{\D})<<1,
\l{a2}\ea
where $C_i=O(1)$ are the unimportant constants. The inequality
is consequence of (\r{ineq1}).

This equations have following solutions:
\be
\D_c\simeq\f{n}{n_{max}\ln n}<<1,~ \ga'_c\sim\ln n>>1.
\l{a3}\ee
Using this solution one can see from (\r{est0}) that it gives
\be
\s_n<O(e^{-n}).
\l{a4}\ee

B. $z_s=+1$: $z_c\to1$, $\D_c<<1$.

One should estimate $\Ga(\ga',\D)$ near the singularity at $z=1$ and
in vicinity of $\D=0$ to consider the consequence of this solution.
Expanding $\Ga(\ga',\D)$ over $\D$ at $\ga'\to0$,
\be
\Ga(\ga',\D)=\Ga(\ga')-\D^{\ga'}e^{-\D}+O(\D^{\ga'+1})
\simeq
\f{1}{\ga'}+O(1).
\l{b1}\ee
This gives following equations for $\ga'$:
\be
n=C_1'\f{\ga'\ln(1/\D)-1}{\ga'}e^{\ga'\ln(1/\D)}.
\l{b2}\ee
The equation for $\D$ has the form:
\be
n_{max}=C_2'e^{(\ga'+1)\ln(1/\D)}.
\l{b3}\ee
Where $C_i'=O(1)$ are unimportant constants.

At
\be
0<\ga'\ln(1/\D)-1<<1,~{\rm i.e.~at}~\ln(1/\D)<<n<<\ln^2(1/\D),
\l{b4}\ee
we find:
\be
\ga'_c\sim \f{1}{\ln(1/\D_c)}.
\l{b5}\ee
Inserting this solution into (\r{b3}):
\be
\D_c\sim\f{1}{n_{max}}.
\l{b6}\ee
It is remarkable that $\D_c$ in the leading approximation is $n$
independent. By this reason $\ga'_c$ becomes $n$ independent also:
\be
\ga'_c\sim \f{1}{\ln(n_{max})}:~
z_c=1+\f{1}{\bar{n}_0^R\ln(n_{max})}.
\l{b7}\ee
This means that
\be
\s_n=O(e^{-n})
\l{b8}\ee
and obey the KNO scaling with mean multiplicity
$\bar{n}=\bar{n}_0^R\ln(n_{max})$.

\section{Conclusion}

Comparing A and B solutions we can see the change of attraction
points with rising $n$: at $n\simeq\bar{n}^2(s)= \bar{n}_0^R
\ln(\sqrt{s}/m_0)$ the transition from (\r{b8}) asymptotics to
(\r{a4}) should be seen. At the same time one should see the strong
KNO scalings violation at the tail of multiplicity distribution.

The comparison of considered above model with experimental data at
moderate and high energies will be given in subsequent papers.

We have neglect the strings interactions and the final state
particles interactions deriving this results. This assumption seems
natural since $z_c-1<<1$ is essential at $\bar{n}<<n<<\bar{n}^2$. By
this reason one can neglect higher powers of $(z_c-1)$ in expansion
of $\ln\x(z,m)$ over $(z_c-1)$. Therefore, describing $\x(x,m)$ we
may restrict ourselves by Poisson distribution (\r{pois}).

At the same time, at first glance, we can not neglect in (\r{fe1})
the contributions with $k>1$ in the moderate region
$\bar{n}<<n<<\bar{n}^2$. Indeed, $k$-th order in (\r{fe1})  is
$$
\sim\Ga^k(\ga'_c,\D_c)\sim(\f{1}{\ga'_c})^k\sim(\ln\D_c)^k\sim(\ln
n_{max})^k\sim(\ln(s/m_0^2))^k>>1
$$
Nevertheless it can be shown that the higher terms with $k>1$ can not
change our semiqualitative conclusion. This question will be
considered later.

\newpage
{\bf Acknowledgments}
\vskip 0.5cm

We are grateful to V.G.Kadyshevski for interest to discussed in the
paper questions. We would like to note with gratitude that the
discussions with E.Kuraev was interesting and important.

\end{document}